\documentclass[12pt]{article}
\usepackage{epsfig}
\begin{document}
\begin{titlepage}
\begin{center}
\hfill{SNUTP-05-013}\\

  \vspace{2cm}

  {\Large \bf  $\eta_c$- Glueball Mixing and Resonance X(1835)}
  \vspace{0.50cm}\\
  Nikolai Kochelev$^{a,b}$\footnote{kochelev@theor.jinr.ru},
  Dong-Pil
Min$^a$ \footnote{dpmin@phya.snu.ac.kr} \vspace{0.50cm}\\
{(a) \it School of Physics and Center for Theoretical Physics,
Seoul National University,
  Seoul 151-747, Korea}\\
\vskip 1ex {(b) \it Bogoliubov Laboratory of Theoretical Physics,
Joint Institute for Nuclear Research, Dubna, Moscow region, 141980
Russia} \vskip 1ex
\end{center}
\vskip 0.5cm \centerline{\bf Abstract} The mixing of $\eta_c$ and
the lowest mass pseudoscalar glueball is estimated within the
framework of the instanton liquid model. It is demonstrated that
the mixing is large and  may explain the difference between the
observed mass of the glueball candidate X(1835) and the
theoretical prediction of QCD sum rule analysis.
  \vspace{1cm}
\end{titlepage}

Recently in \cite{km} we presented our arguments to consider
X(1835) resonance observed by BES Collaboration\cite{BES} in the
reactions $J/\Psi\rightarrow \gamma p\bar p$ and
$J/\Psi\rightarrow \gamma \eta^\prime \pi^+\pi^- $
 as the lowest mass pseudoscalar glueball  \footnote{The possibility of
strong X(1835) coupling to gluons was pointed out by Rosner in
\cite{rosner}. The value of this coupling and X(1835)  decay modes
were   discussed recently   in  \cite{china} as well.}. Our
interpretation is based on the appearance of the parity doublet
structure for high mass hadronic excitations, which can be
explained naturally within the instanton model for QCD vacuum.
Thus, we have considered the doublet $[X(1835),f_0(1710)]$  as the
parity doublet of lowest mass glueballs. Furthermore, the
contribution of X(1835) to the flavor singlet axial vector
coupling of proton and its influence to the proton spin problem
with the large observed coupling of X(1835) to the $p\bar p$
channel were given there.

However, we left unexplained one doubt in interpreting the X(1835)
as the lowest pseudoscalar glueball. That is the magnitude of its
mass, which is lower than the predicted values of the quenched
lattice approach, $2.1\sim 2.5$ GeV \cite{lattice},
 and QCD sum rules,
$2.05\pm 0.19$ GeV \cite{narison}, $2.2\pm 0.2$  GeV
\cite{forkel}.

In this Letter we provide our conjecture that the
$\eta_c$-glueball mixing can be a key factor of adjusting the mass
of lowest pseudoscalar glueball to its experimental value.

The $\eta_c$ has the same quantum numbers as  the X(1835), $(I=0,
J^{PC}=0^{-+})$ and
  the  mass 2.98 GeV
which is quite close to the lattice and QCD sum rule estimations
of the pseudoscalar glueball mass. So the mixing of the $\eta_c$
with glueball can be  large due to the possible $c\bar c$
annihilation  to two gluons. We will estimate this mixing by using
the instanton model  for QCD vacuum \cite{shuryak},
\cite{diakonov}. The effective interaction responsible for the
mixing follows from the gluon-gluon effective interaction induced
by instantons \cite{ABC,shuryak,diakonov}:
\begin{eqnarray}
{\cal L}_{eff}&=&\int dUd\rho n(\rho)
e^{-\frac{2\pi^2}{g_s}\rho^2U_{ab}\bar\eta_{b\alpha\beta}G^a_{\alpha\beta}(x)}
+(I\rightarrow \bar I ), \label{lag}
\end{eqnarray}
where, $n(\rho)$ is the effective instanton density, $\rho$ is the
instanton size, $g_s$ is a strong coupling constant,
$\bar\eta_{b\alpha\beta}$ is the 'tHooft symbol,  and $U$ is the
orientation matrix of the instanton in $SU(3)_c$ color space.
 The  last term in Eq.(\ref{lag}) represents
the contribution coming from anti-instantons. We are going to
calculate the contribution of diagram illustrated in Fig.1 to the
non-diagonal matrix element
\begin{equation}
\frac{\Delta M_{\eta_cG_P}^2}{2}=-<\eta_c|{\cal L}_{eff}|G_P>.
\label{matrix}
\end{equation}
\begin{figure}[htb]
\centering \vspace*{0.5cm}
\psfig{file=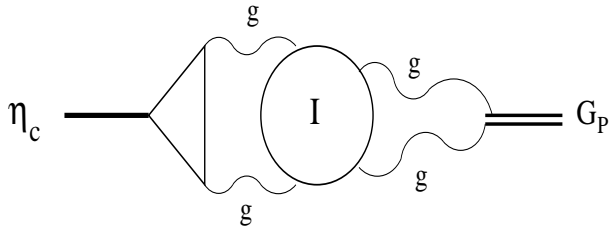,width=8cm,height=3cm,angle=0} \caption{The
$\eta_c$  - pseudoscalar  glueball $G_P$ mixing induced by
instanton. The symbol $I$ denotes the instanton.}
\end{figure}

In the vacuum dominance approximation, which is very suitable in
estimation of instanton contributions to various hadron decays
(see \cite{kv} and \cite{schafer}), the contribution corresponding
the diagram in  Fig.1 to
 the matrix element, Eq.(\ref{matrix}), can be written in the following
form
\begin{equation}
\frac{\Delta M_{\eta_cG_P}^2}{2}=\int d\rho n(\rho) \Big
(\frac{\pi^3\rho^4}{8\alpha_s(\rho) }\Big )^2
<\eta_c|G_{\alpha\beta}^a\widetilde{G}_{\alpha\beta}^a|0><0|G_{\mu\nu}^b\widetilde{G}_{\mu\nu}^b|G_P>.
\label{m1}
\end{equation}
In framework of instanton model of QCD vacuum the matrix element
$<\eta_c|G_{\alpha\beta}^a\widetilde{G}_{\alpha\beta}^a|0>$
 has been calculated in \cite{schafer} following the approach of
\cite{forte}
\begin{equation}
 <\eta_c|G_{\alpha\beta}^a\widetilde{G}_{\alpha\beta}^a|0>=
\frac{64\alpha_s(2m_c)m_c^{3/2}|\Psi(0)|}{\pi^3\rho^4\sqrt{6}}I_{\eta_c}(\rho),
 \label{m2}
\end{equation}
\begin{equation}
I_{\eta_c}(\rho)\simeq
\frac{\pi^2A_0\rho^4log(1+1/(m_c\rho))}{1+B_0(m_c\rho)^4log(1+1/(m_c\rho))},
\end{equation}
where $A_0=0.213$,   $B_0=0.124$. In Eq.(\ref{m2}) the $\Psi(0)$
is the $^1S_0$ wave function of charmonium at the origin. Our main
result is
\begin{equation}
\frac{\Delta M_{\eta_cG_P}^2}{2}\simeq \int d\rho n(\rho)
\frac{m_c^{3/2}\pi^3\rho^4}{\sqrt{6}\alpha^2_s(\rho)}|\Psi(0)|I_{\eta_c}(\rho)
<0|\alpha_s G_{\mu\nu}^b\widetilde{G}_{\mu\nu}^b|G_P>. \label{m3}
\end{equation}
According to the instanton liquid model by Shuryak
\cite{shuryak1}, instanton density is given by
\begin{equation}
n(\rho)=n_0\delta(\rho-\rho_c), \label{den}
\end{equation}
where $n_0\approx 0.5$  fm $^{-4}$ and $\rho_c\approx 1/3$  fm. We
adopt our  parameter values to fit the properties of charmonium,
 $m_c=1.25$  GeV,   $|\Psi(0)|=0.19 $ GeV$^{3/2}$ as in \cite{schafer},
and of
 the strong coupling constant
  at average instanton size
 $\alpha_s(\rho_c)=0.52$ as in \cite{diakonov}.
  The coupling
of X(1835) to gluons   was obtained in  our previous
paper~\cite{km}
\begin{equation}
f_{G_P}=<0|\alpha_sG_{\mu\nu}^b\widetilde{G}_{\mu\nu}^b|G_P>\simeq
2.95\  GeV^3. \label{coupling}
\end{equation}
This value is consistent  with the  result of recent QCD sum rule
analysis $f_{G_P}=2.9\pm 1.4$  GeV$^3$ \cite{forkel}. Our estimate
of the mixing is
\begin{equation}
\Delta M_{\eta_cG_P}^2\simeq 1.54 \  GeV^2. \label{mix2}
\end{equation}

Now we are in the position to evaluate the effect of mixing on the
mass of pseudoscalar glueball by using the following decomposition
of physical  charmonium  and glueball states
\begin{eqnarray}
|\eta_c>&=&|G_P^0>sin\theta+|\eta_c^0>cos\theta\nonumber\\
|G_P>&=&|G_P^0>cos\theta-|\eta_c^0>sin\theta, \label{wf}
\end{eqnarray}
where $|\eta_c^0>$ and $|G_P^0>$ are  bare states. Let us
 assume that  bare masses of glueball and $\eta_c$ are following
\begin{equation}
M_{G_P}^0=2 \  GeV,  \  \   \  M_{\eta_c}^0=2.9 \  GeV.
\label{bare}
\end{equation}
These values lie inside the range of QCD sum rules expectation
\cite{narison}, \cite{forkel}, \cite{heavy}, \cite{zyablyuk} if
one admits about 10\%  accuracy in predictions of this approach
due to uncertainties  in  values of various gluon condensates,
mass of charm quark and $\alpha_s$, high dimension operator
contributions, etc. \footnote{Unfortunately, the available lattice
results
 for the lowest mass pseudoscalar
glueball have been performed  in the quenched approximation
\cite{lattice}. It is rather hard to estimate the accuracy of such
an approximation in the pseudoscalar channel where the light quark
exchange
 between  instantons plays an essential role (see discussion in \cite{km}).}
As the result of the mixing Eq.(\ref{mix2}),  physical masses and
mixing angle are
\begin{eqnarray}
M_{G_P}\simeq 1.87 \  GeV,  \  \   \ \  \  M_{\eta_c}\simeq 2.98 \
GeV,\ \ \ \ \ \theta \simeq 17^\circ. \label{mixingf}
\end{eqnarray}
Therefore, the mixing leads to the increasing of $\eta_c$ mass to
its experimental value and decreasing of the pseudoscalar glueball
mass towards the mass of glueball candidate X(1835). The value of
the mixing angle,  Eq.(\ref{mixingf}), is rather large and should
be taken into account in the calculation of different properties
of $\eta_c$ and pseudoscalar glueball with decay modes. In this
connection we may point out  that  the mixing might be present
behind of the observed  large decay rates of $\eta_c$ to $p\bar p$
and $\eta^\prime\pi\pi$ final states \cite{PRD} due to large
coupling of glueball to these channels.

In principle, fine tuning of parameters  allows us to bring the
mass of glueball to observed one. However, we think that such
procedure is beyond the accuracy of our approach based on the
vacuum dominance approximation and definite instanton model for
QCD vacuum. Furthermore, before the tuning process, some
additional effects such as  the glueball  mixing with
$\eta_c(2S)$, $\eta^\prime$ and others, which are beyond the scope
of the present paper, should be taken into account.

In summary, we have shown that the  instanton induced mixing of
the charmonium and the pseudoscalar glueball is large and may
explain the difference between the  experimental mass of the
glueball candidate X(1835) and the prediction of QCD sum rules.
This observation  provides the additional argument
 in favor of  our suggestion in \cite{km} to treat the  X(1835)
 as  the lowest mass pseudoscalar glueball.

We are  grateful to A.E.~Dorokhov, S.B.~Gerasimov, Xiao-Gang He,
R.N.~Faustov,  H.~Forkel, S.~Narison,
  A.A.~Pivovarov,
San Fu Tuan and Shi-Lin Zhu
   for useful
discussion.
 This work was supported by Brain Pool program of
Korea Research Foundation through KOFST,  grant 042T-1-1,  and in
part by grants of Russian Foundation for Basic Research,
RFBR-03-02-17291 and RFBR-04-02-16445 (NK). NK is very grateful to
the School of Physics, SNU, for their warm hospitality during this
work.

\end{document}